\documentstyle[12pt]{article}
\newcommand{\be}{\begin{equation}}
\newcommand{\ee}{\end{equation}}
\newcommand{\n}{\noindent}
\newcommand{\vs}{\vspace{0.5cm}}
\newcommand{\f}{\frac}
\newcommand{\ba}{\begin{eqnarray}}
\newcommand{\ea}{\end{eqnarray}}
\newcommand{\R}{\mbox{I} \! \mbox{R}}
\newcommand{\N}{\mbox{I} \! \mbox{N}}

 \Large

\begin{document}
\hsize=17truecm

\begin{center}
{\LARGE{\bf On a Rigorous Proof of the  Joos-Zeh
Formula for Decoherence
 in a Two-Body Problem
}\footnote{Proceedings of the Conference ``Multiscale Methods in Quantum
Mechanics: Theory and Experiments'', Roma, December 16-21, 2002.}}

\vspace{2cm}

{\bf Alessandro Teta}\\ {\em Dipartimento di Matematica Pura e
Applicata, Universit\'a di L'Aquila, Italy} \\ {\em e-m: \rm
teta@univaq.it }

\end{center}

\vspace{2cm}

\begin{center}
{\bf Abstract}
\end{center}

\vspace{0.2cm}
\n
{\small We  consider  a simple  one dimensional system consisting of two
particles interacting with a $\delta$-potential and we discuss a
rigorous  derivation  of the asymptotic wave function of the system in
the limit of small mass  ratio.
\n
We apply the result to the explicit computation of the decoherence
effect induced by scattering
 in a concrete example of quantum  evolution.}

\vspace{3cm}

\n
In this note we shall briefly discuss the asymptotic form of the wave
function
of a two-particle system in the limit of small mass ratio.

\n
The interest in this problem is motivated by the analysis initiated by
Joos and  Zeh ([JZ])
of the mechanism of decoherence induced on a heavy particle by the
scattering of light ones (see also
[GF],[T],[BGJKS],[GJKKSZ],[D]).

\n
The basic idea for the analysis of the process is that the small mass
ratio produces a separation of
two characteristic time scales, one slow for the dynamics of the heavy
particle and the other fast for the
light ones.

\n
Following this line, in [JZ] the elementary scattering event between a
light
and a heavy particle is described by the instantaneous transition

\be
\phi(R) \chi (r)\rightarrow \phi (R) (S_{R} \chi )(r)
\label{jz}
\ee

\n
where $\phi$ and $\chi$ are the initial wave functions of the heavy and
the
light particle respectively and $S_{R}$ is the scattering operator of
the
light particle when    the heavy particle is considered fixed in its
initial
position $R$.

\n
In  (\ref{jz}) the initial state is chosen in the form of a
product state, i.e. no correlation is assumed at time zero; moreover
the final state is computed in
 a zero-th order adiabatic  approximation for small values of the  mass
ratio.

\n
Formula (\ref{jz}) gives a simple and clear  description of the
scattering event; nevertheless the approximation chosen is rather
crude in the sense that  time zero of the heavy particle
corresponds to  infinite time of the light one and  the
evolution in time of the  system  is completely negleted.

\n
In order to restore the time evolution,  in
[JZ]   the formula is modified  introducing by hand the
internal dynamics of the heavy particle.

\n
Our aim in this note is to discuss how the complete Joos and Zeh
formula,  i.e. modified taking into account the internal motion of
the heavy particle, can be rigorously derived from the
Schr\"{o}dinger equation of the whole two-particle  system.

\n
In particular, using the result proved in [DFT], we shall write
the asymptotic  form of the wave  function of the system
approximating the true wave function in a specific scaling limit
(involving the mass ratio and the strength of the interaction)
with an explicit control of the error.

\n
Furthermore we shall apply  the result for the analysis of
decoherence in a concrete example of quantum evolution.

\n
More precisely, we shall consider an initial state  with the heavy
 particle in a coherent superposition  of two spatially separated
wave packets with opposite momentum
and the light one localized far on the left with a
positive momentum.

\n
Under precise
assumptions on the relevant physical parameters (spreading and
momentum of the light particle, effective range of the
interaction, spreading and separation of the wave packets of the
heavy particle)
we shall derive an approximated form of the wave function
of the system, describing the typical entangled state with the
wave of the light particle splitted into a reflected and a transmitted
part  by each
wave packet of the heavy particle.

\n
As a
consequence,  the reduced density matrix for the heavy particle
will show unperturbed diagonal terms and
off-diagonal terms reduced by a
factor $\Lambda$  less than one,  which gives a measure of the
decoherence effect  induced on the heavy particle.

\n
Due to a phase shift,  the  waves reflected by the two wave
packets are shown to be approximately orthogonal if the separation of
the two wave packets
is sufficiently large  and then the factor $\Lambda$ can be
explicitely computed in terms of the transmission
probability for the light particle subject to a $\delta$-potential
placed at a fixed position.

\n
We want to emphasize that, at each step, the approximate formulas are
obtained through rather elementary estimates with  explicit control
of the error. In this sense the model may be considered of pedagogical
relevance
for the analysis
of the mechanism of decoherence.

\vs
\n
Let us introduce the model. The hamiltonian of the two-particle
system is given by

\be
H= -\f{\hbar^{2}}{2M} \Delta_{R}   - \f{\hbar^{2}}{2m}
\Delta_{r} + \alpha_{0}
\delta (r-R)
\label{H}
\ee

\n
where $M$ and $m$ denote the mass of the heavy and the light particle
respectively and $\alpha_{0}>0$ is the strength of the interaction.

\n
The hamiltonian (\ref{H}) is a well defined selfadjoint and positive
operator in $L^{2}(\R^{2})$ (see e.g. [AGH-KH]) and moreover it is a
solvable model, in the sense that the corresponding generalized
eigenfunctions and the unitary group can be explicitely computed ([S]).

\n
We consider an initial state given in a product form

\be
\psi_{0}(r,R)= \phi(R) \chi(r), \;\;\;\;\;\;\; \phi, \chi \in {\cal
S}(\R)
\label{statoin}
\ee

\n
where ${\cal S}(\R)$ is the Schwartz space,  and we
denote by $\epsilon = \f{m}{M}$ the mass ratio which is
the small parameter of the model.

\n
We are interested in the asymptotic form of the solution of the
corresponding Schr\"{o}dinger equation

\be
i \hbar \f{\partial \psi^{\epsilon}_{t}}{\partial t} = H
\psi^{\epsilon}_{t}
\label{?}
\ee

\n
when  $\epsilon \rightarrow 0$ and  $\hbar$, $M$ and the
parameter

\be
\alpha= \f{m \alpha_{0}}{\hbar^{2}}
\label{alpha}
\ee

\n
are kept fixed.
Notice that $\alpha^{-1}$ is a length with  the physical  meaning of an
 effective
range of the interaction.

\n
Rescaling the time according to

\be
\tau = \f{\hbar}{M} t
\label{tau}
\ee

\n
the Schr\"{o}dinger equation can be more conveniently written as

\be
i \f{\partial \psi^{\epsilon}_{\tau}}{\partial \tau} = - \frac{1}{2}
\Delta_{R} \,\psi^{\epsilon}_{\tau} \; + \;\frac{1}{\epsilon} \left( -
\frac{1}{2}
\Delta_{r} + \alpha \delta(r-R) \right) \psi^{\epsilon}_{\tau}
\label{es}
\ee

\n
From (\ref{es}) it is clear that for $\epsilon \rightarrow 0$ the
kinetic
energy of the heavy particle can  be considered  as a  small
perturbation.

\n
The  situation is  similar to the Born-Oppenheimer approximation,
with  the relevant difference that here the light particle is not in  a
bound state and so it cannot
produce any effective potential for the heavy particle.

\n
Then, for $\epsilon \rightarrow 0$, one should expect a
scattering regime for the light particle in presence of the heavy one in
a
fixed position and a free motion for the heavy particle.

\n
In fact,  the asymptotics for
$\epsilon \rightarrow 0$ of $\psi^{\epsilon}_{\tau}$
is characterized in the following proposition ([DFT]).

\vs
\n
{\bf Theorem 1}. {\em For any initial state (\ref{statoin}) and any
$\tau>0$  one has

\be
\| \psi^{\epsilon}_{\tau} - \psi^{a}_{\tau} \| < \left( \f{A}{\tau} + B
\right) \epsilon
\ee

\n
where

\be
\psi^{a}_{\tau} (r,R) = \f{1}{\sqrt{i \epsilon^{-1} \tau}}e^{i
\f{\epsilon}{2
\tau} r^{2}} \int dx e^{-i \tau H_0} (R-x) \phi(x) \left[
(\Omega^{x}_{+})^{-1} \chi \right] \tilde{}
\left( \f{ r}{\epsilon^{-1}\tau} \right)
\label{psia}
\ee

\n
and $H_0=-\f{1}{2} \Delta$, the symbol \hspace{0.2cm}$ \tilde{} $
\hspace{0.1cm} denotes the Fourier
transform,
$\Omega^{x}_{+}$ is the wave operator associated to the one-particle
hamiltonian $H_{x} = - \f{1}{2} \Delta  + \alpha \delta(\cdot - x)$, for 
any $x \in \R$; moreover $A,B$ are positive, time-independent constants
whose dependence on the strength of the interaction and on the initial
state are explicitely given.}

\vs
\n
We want to apply the result stated in theorem 1 to the analysis of
decoherence
in a concrete example of quantum evolution. In particular we consider an

initial state $\psi_{0}(R,r) = \phi(R) \chi(r)$ of the following form

\ba
&&\phi(R) = \frac{1}{\sqrt{2}} \left(
f_{\sigma}^{+}(R) + f_{\sigma}^{-}(R) \right) \label{phi}\\
&& f_{\sigma}^{\pm}(R) = \frac{1}{\sqrt{\sigma}} f\left(\frac{R \pm
R_{0}}{\sigma} \right) e^{\pm i P_{0}R}, \;\;\; \sigma,R_{0},P_{0}>0,
\;\;\;\;\;\;\;R_{0}>2 \sigma
\label{f+-}\\
&&\chi(r) = g_{\delta}(r)= \frac{1}{\sqrt{\delta}} g\left(\frac{r
-r_{0}}
{\delta}
\right) e^{ i q_{0}r}, \;\;\;\;\;\delta,q_{0}>0, \;\; r_0 < -R_0 -
\sigma-
\delta \label{g}\\
&&f,g \in C_{0}^{\infty}(-1,1),  \;\;\;\;\; \|f\|=\|g\|=1
\label{smooth}
\ea

\n
From (\ref{phi}),(\ref{f+-}) one sees that the heavy particle is  in a
superposition state of two spatially separated wave packets, one
localized
in $R=-R_0$ with mean value of the momentum $P_0$ and the other
localized
in $R=R_0$ with mean value of the momentum $-P_0$.

\n
From (\ref{g}), the light particle is assumed localized around $r_0$, on 
the left of the wave packet $f_{\sigma}^{+}$, with positive mean
momentum
$q_0$.

\n
Moreover, to simplify the computation, $f^{+}_{\sigma}$,
$f^{-}_{\sigma}$, $g_{\delta}$ are chosen compactly supported and
with disjoint supports.

\n
We expect that in the time evolution of the above initial state, for
$\epsilon \rightarrow 0$, the light particle will be partly reflected
and partly
transmitted by the two wave packets $f_{\sigma}^{+}$, $f_{\sigma}^{-}$,
which approximately act as fixed scattering centers.

\n
We shall make precise this statement introducing suitable
assumptions on the physical parameters of the system.

\n
In order to formulate the result,  we introduce the reflection and
transmission coefficient  associated to the
one-particle hamiltonian $H_x$ (see e.g. [AGH-KH])

\be
{\cal R}_{\alpha}(k) = -\f{\alpha}{\alpha - i |k|}, \;\;\;\;\;\; {\cal
T}_{\alpha}(k) = - \f{i|k|}{\alpha -i|k|}
\label{RT}
\ee

\n
and define the transmitted and reflected part of the wave function of
the
light particle

\ba
&&g_{{\cal T}} (\epsilon^{-1} \tau, r) = {\cal T}_{\alpha}\left(
\f{|r|}{\epsilon^{-1} \tau} \right) \f{e^{i \f{r2}{2 \epsilon^{-1}
\tau}}}{\sqrt{i\epsilon^{-1} \tau}} \; \tilde{g}_{\delta}
\left(\f{r}{\epsilon^{-1} \tau} \right) \\
&&g_{{\cal R}}^{x} (\epsilon^{-1} \tau, r) = {\cal R}_{\alpha}\left(
\f{|r|}{\epsilon^{-1} \tau} \right) e^{-2ix\f{r}{\epsilon^{-1} \tau}}\;
\f{e^{i
\f{r2}{2
\epsilon^{-1}
\tau}}}{\sqrt{i\epsilon^{-1} \tau}}  \; \tilde{g}_{\delta}
\left(-\f{r}{\epsilon^{-1} \tau} \right)
\label{gR}
\ea

\n

\n
Moreover, we shall assume

\be
q_{0} \gg \f{1}{\delta}, \;\;\;\;\;\;\;\; \f{1}{\alpha} \gg \sigma
\label{A1}
\ee

\n
The first assumption in (\ref{A1}) simply means that the wave function
of the light particle is well concentrated in momentum space around
$q_{0}$ (notice that $\delta^{-1}$ is the order of magnitude of the
spreading of the momentum).

\n
The second assumption in (\ref{A1}) requires that the effective range
of the interaction is much larger than the spreading in position of
each wave packet of the heavy particle and this means that the
interaction cannot "distinguish" two different points in the supports
of $f^{+}_{\sigma}$ and $f^{-}_{\sigma}$.

\n
Then

\vs
\n
{\bf Proposition 2}. {\em For the initial state (\ref{phi}),
(\ref{f+-}),
(\ref{g}), (\ref{smooth}) and any $\tau >0$ one has

\ba
&&\|\psi_{\tau}^{a} - \psi_{\tau}^{e} \| < c \left[ (\delta q_0)^{-n} +
\sigma \alpha  \right]
\ea

\n
for any $n\in \N$, where $c$ is a numerical constant depending on $g$
and }

\ba
&&\psi_{\tau}^{e}(r,R) = \f{1}{\sqrt{2}} \left( e^{-i \tau H_{0}}
f^{+}_{\sigma} \right) (R) \left[  g_{{\cal T}} (\epsilon^{-1} \tau,
r) + g_{{\cal R}}^{-R_{0}} (\epsilon^{-1} \tau, r) \right]
\nonumber\\
&&
+ \f{1}{\sqrt{2}} \left( e^{-i \tau H_{0}}
f^{-}_{\sigma} \right) (R) \left[  g_{{\cal T}} (\epsilon^{-1} \tau,
r) + g_{{\cal R}}^{R_{0}} (\epsilon^{-1} \tau, r) \right]
\label{psie}
\ea

\vs
\n
{\bf Proof.} Exploiting the explicit expression of the generalized
eigenfunctions of $H_{x}$ (see e.g. [AGH-KH]), for $x \in \; supp
\;f^{\pm}_{\sigma}$, one has

\ba
&&[(\Omega^{x}_{+})^{-1} g_{\delta}]\, \tilde{} \,(k) = \f{1}{\sqrt{2
\pi}} \int dy g_{\delta} (y) \left( e^{-i k y} + {\cal R}_{\alpha}(k)
e^{-ikx} e^{i |k||x-y|} \right)\nonumber\\
&&= {\cal T}_{\alpha}(k) \tilde{g}_{\delta}(k) \theta_{+}(k) +
\left[ \tilde{g}_{\delta}(k) + e^{-2ikx} {\cal R}_{\alpha}(k)
\tilde{g}_{\delta}(-k)  \right] \theta_{-}(k)
\ea

\n
where we have used the fact that $|x-y|=x-y$ for $x \in \,supp \,
f^{\pm}_{\sigma}$, $y \in \, supp \, g_{\delta}$, the identity $1 +
{\cal R}_{\alpha} = {\cal T}_{\alpha}$ and we have denoted by
$\theta_{+}$, $\theta_{-}$ the characteristic functions of the positive
and negative semiaxis respectively.

\n
From (\ref{psia}),(\ref{phi}),(\ref{f+-}),(\ref{g}),(\ref{psie}) we can
write

\ba
&&\psi^{a}_{\tau}(r,R) - \psi^{e}_{\tau}(r,R)\nonumber\\
&&= \f{1}{\sqrt{2i \epsilon^{-1} \tau}} e^{i \f{\epsilon}{2\tau} r^{2}}
\int  dx e^{-i \tau H_{0}} (R-x) (f^{+}_{\sigma}(x) +f^{-}_{\sigma}(x))
{\cal R}_{\alpha} \left(\f{r}{\epsilon^{-1} \tau}
\right) \left[
\tilde{g}_{\delta} \left(\f{r}{\epsilon^{-1} \tau}
\right)
\theta_{-}(r) \right. \nonumber\\
&&\left.  - e^{-2ix\f{r}{\epsilon^{-1}
\tau}}
\tilde{g}_{\delta} \left(-\f{r}{\epsilon^{-1} \tau}
\right)
\theta_{+}(r) \right]
\nonumber\\
&&+ \f{1}{\sqrt{2}} \int dx e^{-i \tau H_{0}} (R-x)
f^{+}_{\sigma}(x)
\left( g^{x}_{{\cal R}} \left(\f{r}{\epsilon^{-1} \tau}
\right) -  g^{-R_{0}}_{{\cal R}} \left(\f{r}{\epsilon^{-1} \tau}
\right) \right)\nonumber\\
&&+ \f{1}{\sqrt{2}} \int dx e^{-i \tau H_{0}} (R-x)
f^{-}_{\sigma}(x)
\left( g^{x}_{{\cal R}} \left(\f{r}{\epsilon^{-1} \tau}
\right) - g^{R_{0}}_{{\cal R}} \left(\f{r}{\epsilon^{-1} \tau}
\right) \right)\nonumber\\
&&\equiv (\zeta_{1}+\zeta_{2}+\zeta_{3})(r,R)
\ea

\n
Noticing that $\tilde{g}_{\delta} (k)=e^{i(q_{0}-k)r_0} \sqrt{\delta}
\tilde{g}(\delta
k - \delta q_{0})$, one has

\ba
&&\| \zeta_{1}\|^{2} = \f{1}{2 } \int dx
|f^{+}_{\sigma}(x) + f^{-}_{\sigma}(x)|^{2} \int dk \left| {\cal
R}_{\alpha} \left( k \right) \right|^{2}
\left|
\tilde{g}_{\delta} \left(k\right)
\theta_{-}(k)  - e^{-2ix k}
\tilde{g}_{\delta} \left(-k \right)
\theta_{+}(k)
\right|^{2}\nonumber\\
&&\leq \int dk \left( |\tilde{g}_{\delta} \left(k\right)
\theta_{-}(k)|^{2} + |\tilde{g}_{\delta} \left(-k\right)
\theta_{+}(k)|^{2} \right) \nonumber\\
&&= 2\int_{\delta q_{0}}^{\infty} dz |\tilde{g}(z)|^{2}
\label{stima1}
\ea

\n
which is $O\left( (\delta q_{0})^{-n}\right)$ for any $n \in \N$.
Concerning $\zeta_{2}$ we have

\ba
&&\|\zeta_{2}\|^{2}= \f{1}{2} \int dx |f^{+}_{\sigma}(x)|^{2}
\int dk |{\cal R}_{\alpha}(k)|^{2} |\tilde{g}_{\delta}(k)|^{2}
\left| e^{-2ikx} - e^{2ikR_{0}} \right|^{2}\nonumber\\
&&\leq 2 \sup_{x \in (-R_{0}- \sigma, -R_{0}+ \sigma)} |x+R_{0}|^{2}
  \int dk k^{2}
|{\cal R}_{\alpha}(k)|^{2} |\tilde{g}_{\delta}(k)|^{2}\nonumber\\
&&\leq 2 \sigma^{2} \alpha^{2}
\label{stima2}
\ea

\n
where we have used the inequality $k^{2}|{\cal R}_{\alpha}(k)|^{2} \leq
\alpha^{2}$.  The estimate for $\zeta_{3}$ proceeds exactly in the same
way
concluding the proof. $\Box$

\vs
\n
Proposition 2 shows that, under the assumptions (\ref{A1}), the wave
function (\ref{psie}) of the whole system takes the form of the typical
entangled
state emerging from an interaction of von Neumann type between a
system and an apparatus. Here the role of the apparatus is played by
the light particle which is partly transmitted and partly reflected by
each wave packet of the heavy particle.

\n
In this process  the light particle keeps the information
about the position of the heavy particle,
which is encoded in $g_{{\cal R}}^{-R_{0}}$ and $g_{{\cal R}}^{R_{0}}$,
i.e. in
the reflected waves by $f^{+}_{\sigma}$ and $f^{-}_{\sigma}$
respectively.

\n
As it should be expected in a one dimensional scattering process,
 such reflected waves have an identical spatial
localization and  they only differ for a phase factor.

\n
In order to discuss the decoherence effect induced on the heavy
particle one has to consider the density matrix of the whole system
$\rho^{\epsilon}_{\tau}(r,R,r',R')$ and then one
should compute the reduced density matrix for the heavy particle

\be
\hat{\rho}^{\epsilon}_{\tau}(R,R') \equiv \int dr
\rho^{\epsilon}_{\tau}(r,R,r,R') \equiv \int dr
\psi^{\epsilon}_{\tau}(r,R)
\overline{\psi^{\epsilon}_{\tau}}(r,R')
\ee

\n
which defines a positive, trace-class operator in $L^{2}(\R^{2})$
with $Tr\; \hat{\rho}^{\epsilon}_{\tau} = 1$.

\n
We also introduce the approximate reduced density matrix

\ba
&&\hat{\rho}^{e}_{\tau}(R,R') \equiv \int dr
\rho^{e}_{\tau}(r,R,r,R') \equiv  \int dr \psi^{e}_{\tau}(r,R)
\overline{\psi^{e}_{\tau}}(r,R')\nonumber\\
&&= \f{1}{2} \left( e^{-i \tau H_{0}} f^{+}_{\sigma}\right)(R)
\left( e^{i \tau H_{0}} \overline{f^{+}_{\sigma}}\right)(R') +
\f{1}{2} \left( e^{-i \tau H_{0}} f^{-}_{\sigma}\right)(R)
\left( e^{i \tau H_{0}} \overline{f^{-}_{\sigma}}\right)(R') \nonumber\\
&&+\f{\Lambda}{2} \left( e^{-i \tau H_{0}} f^{+}_{\sigma}\right)(R)
\left( e^{i \tau H_{0}} \overline{f^{-}_{\sigma}}\right)(R') +
\f{\bar{\Lambda}}{2} \left( e^{-i \tau H_{0}} f^{-}_{\sigma}\right)(R)
\left( e^{i \tau H_{0}} \overline{f^{+}_{\sigma}}\right)(R')
\label{rhoe}
\ea

\n
where, in the last equality, we took into account that

\be
 \int dr |g_{{\cal T}} (\epsilon^{-1} \tau,
r) + g_{{\cal R}}^{-R_{0}} (\epsilon^{-1} \tau, r)|^{2} =
\int dr |  g_{{\cal T}} (\epsilon^{-1} \tau,
r) + g_{{\cal R}}^{R_{0}} (\epsilon^{-1} \tau, r)|^{2}=1
\ee

\n
and we defined

\be
\Lambda \equiv \int dr \left( g_{{\cal T}} (\epsilon^{-1} \tau,
r) + g_{{\cal R}}^{-R_{0}} (\epsilon^{-1} \tau, r) \right)
\left( \overline{g_{{\cal T}}} (\epsilon^{-1} \tau,
r) + \overline{g_{{\cal R}}^{R_{0}}} (\epsilon^{-1} \tau, r) \right)
\label{lambda}
\ee

\n
Notice that $|\Lambda| \leq 1$ and the case $\Lambda =1$ can only
occurs if $\alpha =0$, i.e. when the interaction is absent.

\n
Since $L^{2}$-convergence of the wave function implies convergence of
the corresponding density matrix in the trace-class norm,
we conclude that $\hat{\rho}^{e}_{\tau}$ is a good approximation in
the trace-class norm of $\hat{\rho}^{\epsilon}_{\tau}$ under the
assumptions of theorem 1 and proposition 2.

\n
For $\alpha =0$ we see from (\ref{rhoe}), (\ref{lambda})  that
$\hat{\rho}^{e}_{\tau}$ reduces to the pure state describing the
coherent superposition of the two wave packets evolving in time
according to the free hamiltonian.

\n
Notice that the last two terms in (\ref{rhoe}), usually called non
diagonal terms, are responsible for the interference effects observed
when the two wave packets have a non empty overlapping.

\n
The occurrence of the interference makes the state highly non
classical and it is the distinctive character of a quantum system
with respect to a classical one.

\n
If we switch on the interaction, i.e. for $\alpha >0$, the effect of
the light particle is to reduce the non diagonal terms by the
factor $\Lambda$ and this means that the interference effects for the
heavy particle are correspondingly reduced.

\n
In this sense one can say that there is a (partial) decoherence effect
induced
on the heavy particle which is completely characterized by the
parameter $\Lambda$.

\n
In the limit $\Lambda \rightarrow 0$, i.e. for $\alpha \rightarrow
\infty$, the non diagonal terms and the interference effects are
completely cancelled and then the state becomes a classical
statistical mixture of the two pure states $ e^{-i \tau H_{0}}
f^{+}_{\sigma}$, $ e^{-i \tau H_{0}} f^{-}_{\sigma}$.

\n
We conclude this note showing that the parameter $\Lambda$ can be
approximated by a simpler expression under further conditions on the
parameters.

\n
First we notice that using the first assumption in (\ref{A1}) one
easily gets

\be
\Lambda = \int dr |  g_{{\cal T}} (\epsilon^{-1} \tau,
r) |^{2} + \int dr   g_{{\cal R}}^{-R_{0}} (\epsilon^{-1} \tau, r)
\overline{g_{{\cal R}}^{R_{0}}} (\epsilon^{-1} \tau, r)
+ O((\delta q_{0})^{-n})
\label{lambda1}
\ee

\n
Moreover we can  show that for a large separation of the two
wave packets the second integral in (\ref{lambda1}) becomes negligible
due to the rapid oscillations of the integrand. More precisely we
assume

\be
d \gg \f{1}{\alpha}, \;\;\;\;\;\;\;\;\;\; d \gg \delta
,\;\;\;\;\;\;\;\;\;\;  d
\equiv 2R_{0}
\label{A2}
\ee

\n
From (\ref{gR}), (\ref{RT}) we have

\ba
&&{\cal I} \equiv \int dr   g_{{\cal R}}^{-R_{0}} (\epsilon^{-1} \tau,
r)
\overline{g_{{\cal R}}^{R_{0}}} (\epsilon^{-1} \tau, r)\nonumber\\
&&=\int dk |{\cal R}_{\alpha}(k)|^{2} |\tilde{g}_{\delta}(-k)|^{2}
e^{2idk}= \int dz \f{\alpha^{2}}{\alpha^{2} + (\delta^{-1}z -
q_{0})^{2}}
|\tilde{g}(z)|^{2}e^{2i d(\delta^{-1}z -q_{0})}
\label{calI}
\ea

\n
Integrating by parts in (\ref{calI}) one obtains

\ba
&&|{\cal I}| = \left| \int dz \f{\alpha^{2}}{\alpha^{2} + (\delta^{-1}z
- q_{0})^{2}}
|\tilde{g}(z)|^{2} \f{1}{2i d \delta^{-1}} \f{d}{dz} e^{2i
d(\delta^{-1}z
-q_{0})} \right| \nonumber\\
&&\leq  \f{\delta}{2d}  \int dz \left| \f{d}{dz}
\left( \f{\alpha^{2}}{\alpha^{2} + (\delta^{-1}z - q_{0})^{2}}
|\tilde{g}(z)|^{2} \right) \right|\nonumber\\
&&\leq \f{\delta}{d} \int dz \f{\alpha^{2} \delta^{-1} |\delta^{-1}
z - q_{0}|}{\left[ \alpha^{2} + (\delta^{-1}z
-q_{0})^{2}\right]^{2}} |\tilde{g}(z)|^{2} + \f{\delta}{d} \int dz
\f{\alpha^{2}}{\alpha^{2} + (\delta^{-1}z
-q_{0})^{2}}|\tilde{g}(z)||\tilde{g}'(z)|\nonumber\\
&&\leq \f{1}{d \alpha}+ \f{\delta}{d}\;\|\tilde{g}' \|
\label{stimaI}
\ea

\n
where, in the last line, we have used the trivial estimates

\be
\f{\alpha^{2} \delta^{-1} |\xi|}{(\alpha^{2} + \xi^{2})^{2}} \leq
\f{\alpha^{2}}{\alpha^{2} + \xi^{2}} \f{\delta^{-1}|\xi|}{\alpha^{2} +
\xi^{2}}
\leq \f{\delta^{-1}|\xi|}{\alpha^{2} + \xi^{2}} \leq \f{1}{\delta
\alpha}
\ee

\n
and the
Schwartz inequality.

\n
Using the estimate (\ref{stimaI}) in  (\ref{lambda1}) we conclude that

\be
\left| \Lambda - \int dr |  g_{{\cal T}} (\epsilon^{-1} \tau,
r) |^{2} \right| \leq   \f{1}{d \alpha}+ \f{\delta}{d}\; \|\tilde{g}'
\| +O((\delta q_{0})^{-n})
\ee

\n
This means that, under the assumptions (\ref{A1}),(\ref{A2}),
the decoherence effect is measured by the transmission probability of
the light particle which is explicitely given by

\be
\int dr |  g_{{\cal T}} (\epsilon^{-1} \tau,
r) |^{2} = \int dk |{\cal T}_{\alpha}(k) |^{2}
|\tilde{g}_{\delta}(k)|^{2}
\ee

\n
The previous analysis of decoherence induced by scattering is clearly
 limited  by the consideration of a two-body system.

\n
For a  more satisfactory treatment one should consider a model with $N$
light particles scattered by the heavy one. The expected result is
that the effect of the scattering events is cumulative and then the
decoherence effect is increased at each step ([JZ]).

\n
We observe that a rigorous
proof of this fact would  require good  estimates for the wave operator
in
a case of  many-body scattering problem and it is a non trivial
open question.

    \vspace{2cm}

    {\bf References}
\vs

    \n
    [AGH-KH] Albeverio S., Gesztesy F., Hoegh-Krohn R., Holden H.,
    {\em Solvable Models in Quantum Mechanics}, Springer-Verlag, 1988.

    \n
    [BGJKS]  Blanchard Ph., Giulini D., Joos E., Kiefer C., Stamatescu
    I.-O. eds.,{\em Decoherence: Theoretical, Experimental and
    Conceptual Problems}, Lect. Notes in Phys. 538, Springer, 2000.

\n
[D] Dell'Antonio G., On Decoherence, preprint 2003.

\n
[DFT] Duerr D., Figari R., Teta A., Decoherence in a Two-Particle Model, 
mp arc 02-442, submitted to {\em J. Math. Phys.}.





    \n
    [GF] Gallis M.R., Fleming G.N., Environmental and Spontaneous
    Localization, {\em Phys. Rev.} {\bf A42}, 38-48 (1990).


    \n
    [GJKKSZ] Giulini D., Joos E., Kiefer C., Kupsch J., Stamatescu
    I.-O., Zeh H.D., {\em Decoherence and the Appearance of a
    Classical World in Quantum Theory}, Springer, 1996.


    \n
    [JZ] Joos E., Zeh H.D., The Emergence of Classical Properties
    Through Interaction with the Environment, {\em Z. Phys.} {\bf
    B59}, 223-243 (1985).



    \n
    [S] Schulman L.S., Application of the propagator for the delta
    function potential, in {\em Path Integrals from mev to Mev},
    Gutzwiller M.C., Ioumata A., Klauder J.K., Streit L. eds., World
    Scientific, 1986, pp. 302-311.

    \n
    [T] Tegmark M., Apparent Wave Function Collapse Caused by
    Scattering, {\em Found. Phys. Lett.} {\bf 6}, 571-590 (1993).

\end{document}